\documentclass[12pt]{iopart}
\pdfoutput=1
\expandafter\let\csname equation*\endcsname\relax
\expandafter\let\csname endequation*\endcsname\relax
\usepackage[numbers,sort&compress]{natbib}
\usepackage{amsmath,amssymb,eucal,graphicx,bm}
\usepackage{subfigure}
\usepackage{float}

\def\ltwid{\mathrel{\raise.3ex\hbox{$<$\kern-.75em\lower1ex\hbox{$\sim$}}}}

\begin{document}
\title{Where Should You Park Your Car? The $\frac{1}{2}$ Rule}

\author{P. L. Krapivsky}
\address{Department of Physics, Boston University, Boston, MA, 02215 USA}
\author{S. Redner}
\address{Santa Fe Institute, 1399 Hyde Park Road, Santa Fe, NM, 87501 USA}

\begin{abstract}
  We investigate parking in a one-dimensional lot, where cars enter at a rate
  $\lambda$ and each attempts to park close to a target at the origin.
  Parked cars also depart at rate 1.  An entering driver cannot see beyond
  the parked cars for more desirable open spots.  We analyze a class of
  strategies in which a driver ignores open spots beyond $\tau L$, where
  $\tau$ is a risk threshold and $L$ is the location of the most distant
  parked car, and attempts to park at the first available spot encountered
  closer than $\tau L$.  When all drivers use this strategy, the probability
  to park at the best available spot is maximal when $\tau=\frac{1}{2}$, and
  parking at the best available spot occurs with probability $\frac{1}{4}$.
\end{abstract}

\section{Introduction and Model}

One of the downsides of driving to a popular destination is that good parking
spaces near the venue may be hard to find.  The dilemma is whether to park
far away, which should be easy, and then have a long walk to the destination,
or drive close to the venue and then look for a good parking spot, which is
likely to be hard.  In this work, we investigate a simple class of
threshold-driven parking strategies in an idealized parking lot geometry.  We
find the strategy that maximizes the probability to park in the best
available spot.

Because of its pervasive role in our daily lives, understanding parking has
been the focus of much study, especially in the urban planning and
transportation engineering literatures (see, e.g.,
\cite{V82,YTT91,AP91,TR98,AR99,TL06,KLW14} and references therein).  To help
in making policy decisions, these investigations typically include many
real-world effects, such as practical parking-lot geometries, parking costs,
parking limits, and urban planning implications.  It is difficult to gain
fundamental insights from such reality based approaches.  Our approach is to
investigate a parsimonious model that captures some key features of parking,
and for which some analytic understanding can be gleaned.

\begin{figure}[ht]
  \centerline{\includegraphics[width=0.85\textwidth]{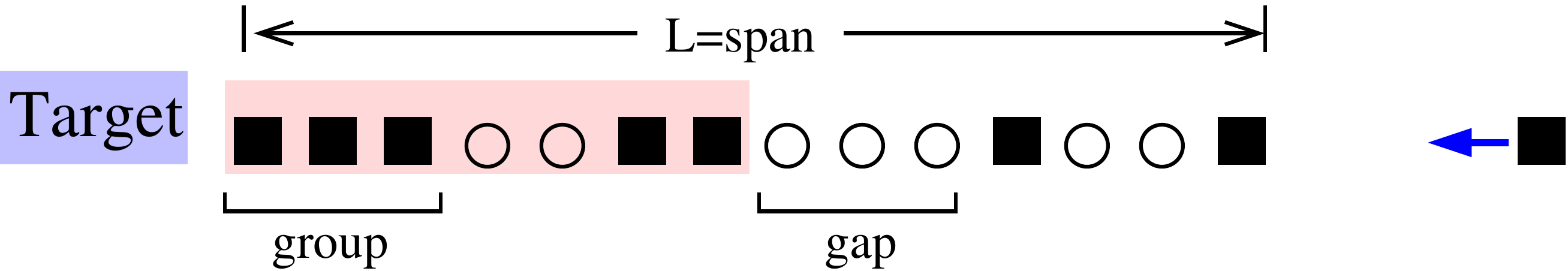}}
  \caption{Parking in a one-dimensional lot where cars (squares) enter from
    the right and circles represent empty spots.  Spots beyond the furthest
    car are not shown.  The next car to enter (blue arrow) will start looking
    for a parking spot in the active zone (red shaded area).  In this
    example, this next entering car parks a distance $k=4$ from the target,
    which happens to be the best available spot.  Here $N=7$, $L=14$, $V=7$,
    and $\tau=\frac{1}{2}$.}
\label{Fig:cartoon-threshold}  
\end{figure}

For simplicity, the parking lot is one-dimensional, with parking spots
labeled by the positive integers $k=1,2,\ldots$ and the desired target is
located at $k=0$ (Fig.~\ref{Fig:cartoon-threshold}).  Cars enter one at a
time from the right at rate $\lambda$ and each car also departs at rate 1;
this input rate is the only parameter of the system.  The dynamics of the
number of cars is governed by the Poisson process, independent of the parking
strategy.  However, the spatial distribution of parked cars depends on the
parking strategy.

We postulate that a driver cannot see open parking spots closer to
the target than the current position of the driver (if there is a contiguous gap of
open spots at the current position, the driver can see only to the end of
this gap).  We additionally assume that after entering the lot, the driver
finds a parking space before the next car enters or any of the parked cars
depart.  While this model is highly idealized, it captures the tradeoff
involved in looking for open spots in a crowded parking lot.

The parked cars form groups that are interspersed by gaps
(Fig.~\ref{Fig:cartoon-threshold}).  Two basic macroscopic observables are
the total number of parked cars $N$ and the span $L$, defined as the distance
between the target and the most distant parked car.  We are particularly
interested in the number of open parking spots within the span (henceforth
referred to as vacancies) $V=L-N$, as well as their spatial distribution.
While the state of the parking lot is always changing, due to the stochastic
arrival and departure of cars, the state of the lot is statistically
stationary.  Moreover, relative fluctuations become small in the
$\lambda\to \infty$ limit.  From the underlying Poisson process for the number
of parked cars, $\langle N\rangle =\lambda$ and
$\langle N^2\rangle -\langle N\rangle^2=\lambda$, so the relative standard
deviation vanishes as $\lambda^{-1/2}$.  In real life, parking lots are often
nearly full, which corresponds to large $\lambda$ in our model.  Hence we
assume $\lambda\gg 1$ unless stated otherwise.

All drivers follow the same parking strategy and we seek the optimal
strategy, which we define as the one that maximizes the probability to find
the best available spot without backtracking, i.e., the vacancy closest to
the target.  (If backtracking occurs, the filled vacancy is closest to the
target, but there is the added expense of backtracking.)~ We analyze a class
of \emph{threshold} parking strategies that are inspired by threshold
strategies that arise in a wide variety of optimal decision problems~
\cite{L61,D70,Chow71,BG84,P99}.  The classic ``secretary
problem''~\cite{D70,Chow71,BG84,P99,Dynkin,Chow64,Gilbert,F89,Bruss00,Bruss03,Dendievel,GKMN,FW10,GM13}
is perhaps the best-known example of the utility of a threshold rule.  In the
framework of our parking problem, the threshold characterizes the degree of
risk that an entering driver is willing to accept.  We thus define a risk
threshold $\tau\in (0,1)$ as follows: When a car enters the parking lot whose
current span is $L$, the driver ignores all vacancies in $[\tau L,L)$ and
only starts looking for vacancies upon reaching a distance $\tau L$ from the
target.  The driver parks at the end of the first eligible gap encountered
that is closest to the target (Fig.~\ref{Fig:cartoon-threshold}).  If no gaps
are found when the target is reached, the driver backtracks and parks in the
first available vacancy, whose location is necessarily greater than $\tau L$.
We call the zone $(0,\tau L)$ active since drivers actively look
for vacancies there; in the complementary passive zone $[\tau L,L)$,
vacancies are initially ignored.

We shall show that in the $\lambda\to\infty$ limit, the probability for a
driver to park in the best available spot is maximized by choosing
$\tau=\frac{1}{2}$.  Normally one does not park at the first vacancy
encountered because one feels that better spots will be available.
But waiting too long may result in the failure to find an available spot. 
The {\bf $\mathbf{\frac{1}{2}}$ Rule} provides the best
compromise between actually finding a spot and not being a sucker for parking
too far away.  For the $\tau$-threshold strategy, we will show that the
probability to find a single vacancy in the active zone, which is the same as
finding the best available parking spot, is $P_1(\tau)=\tau(1-\tau)$.  This
probability is maximized when $\tau=\frac{1}{2}$ and the maximum value of
$P_1(\tau)$ equals $\frac{1}{4}$.

We previously studied parking in the same one-dimensional geometry~\cite{KR}, 
in which parking occurred according to either 
the \emph{optimistic} strategy or to the \emph{prudent} strategy.  In the
optimistic strategy, the driver goes all the way to the target and then
backtracks to the closest available spot.  In the prudent strategy, the
driver parks at the first gap encountered; when there are no vacancies, the
driver backtracks and parks behind the rightmost parked car.  These two rules
correspond to the risk threshold $\tau=0$ and $\tau=1$, respectively.  We
will see that these two extremal parking rules are both inferior to the
strategy with $\tau=\frac{1}{2}$.

In the next section, we present simulation data for the spatial distribution
of cars and the complementary distribution of vacancies.  The latter
distribution has a rich spatial structure and we offer conjectures for the
vacancy density in the active and passive zones that appear to be exact.  In
Sect.~\ref{sec:der} we derive the distribution of the number of vacancies
$P_n(\tau)$ in the active zone.  In Sect.~\ref{sec:position} we argue that
the position of the chosen parking spot is spatially uniform, independent of
the threshold $\tau$.  In Sect.~\ref{sec:cost} we determine the cost
associated with parking and the parking strategy that minimizes this cost.
Finally, in Sect.~\ref{sec:disc} we summarize our results and give some
perspectives.

\section{Spatial Distributions of Parked Cars and Vacancies}
\label{sec:density}

\begin{figure}[ht]
  \centerline{
    \subfigure[]{\includegraphics[width=0.425\textwidth]{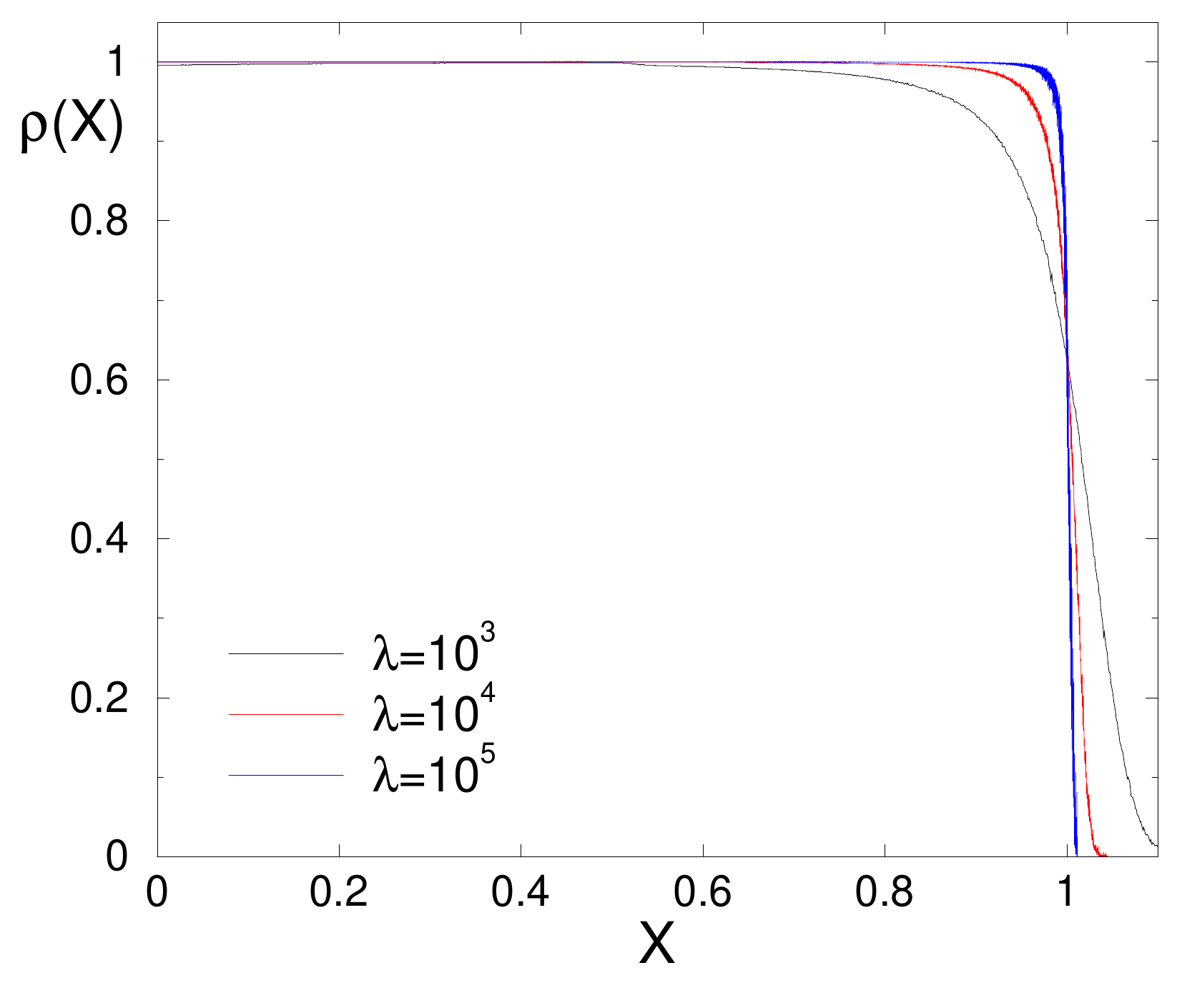}}\qquad\qquad
    \subfigure[]{\includegraphics[width=0.425\textwidth]{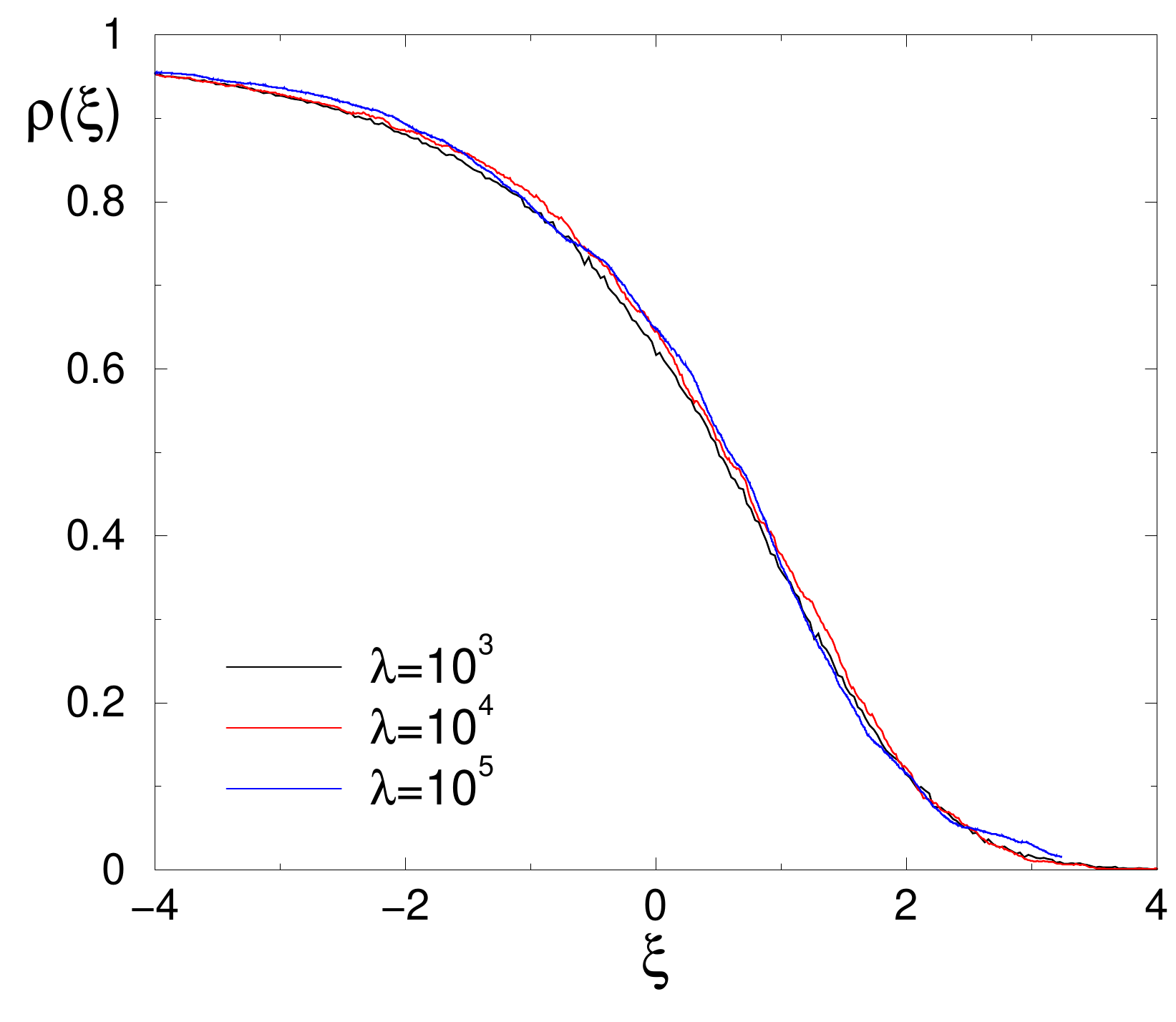}}}
  \caption{(a) Simulation data for the density profile of parked cars versus
    $X=k/\lambda$. (b) The density profile in the front region versus the
    boundary-layer variable $\xi=(X-1)\lambda^{1/2}$.  In (b), the data for
    $\lambda=10^4$ and $\lambda=10^5$ are smoothed over 10-point and
    100-point ranges, respectively.  For both plots $\tau=\frac{1}{2}$. }
\label{fig:rho-vs-x}  
\end{figure}

For the threshold rule with $\tau$ strictly less than 1, the average density
of parked cars at position $k$ approaches a step function as the arrival rate
$\lambda\to\infty$: $\rho(k)= 1$ for $k<\lambda$ and $\rho(k)= 0$ for
$k>\lambda$ (Fig.~\ref{fig:rho-vs-x}(a)).  That is, the bulk of the parking
lot is full and most vacancies are near the far end of the lot where
$k\approx\lambda$.  Figure~\ref{fig:rho-vs-x}(b) shows the density profile
near the top of the ``Fermi sea'' plotted versus the appropriate
`boundary-layer' variable $\xi=(k-\lambda)/\lambda^{1/2}$.  The deviation of
the spatial distribution from a step function vanishes as $\lambda^{-1/2}$
for $\lambda\to\infty$ and the data for other values of $\tau$ are
qualitatively similar.

\begin{figure}[ht]
  \centerline{
    \subfigure[]{\includegraphics[width=0.425\textwidth]{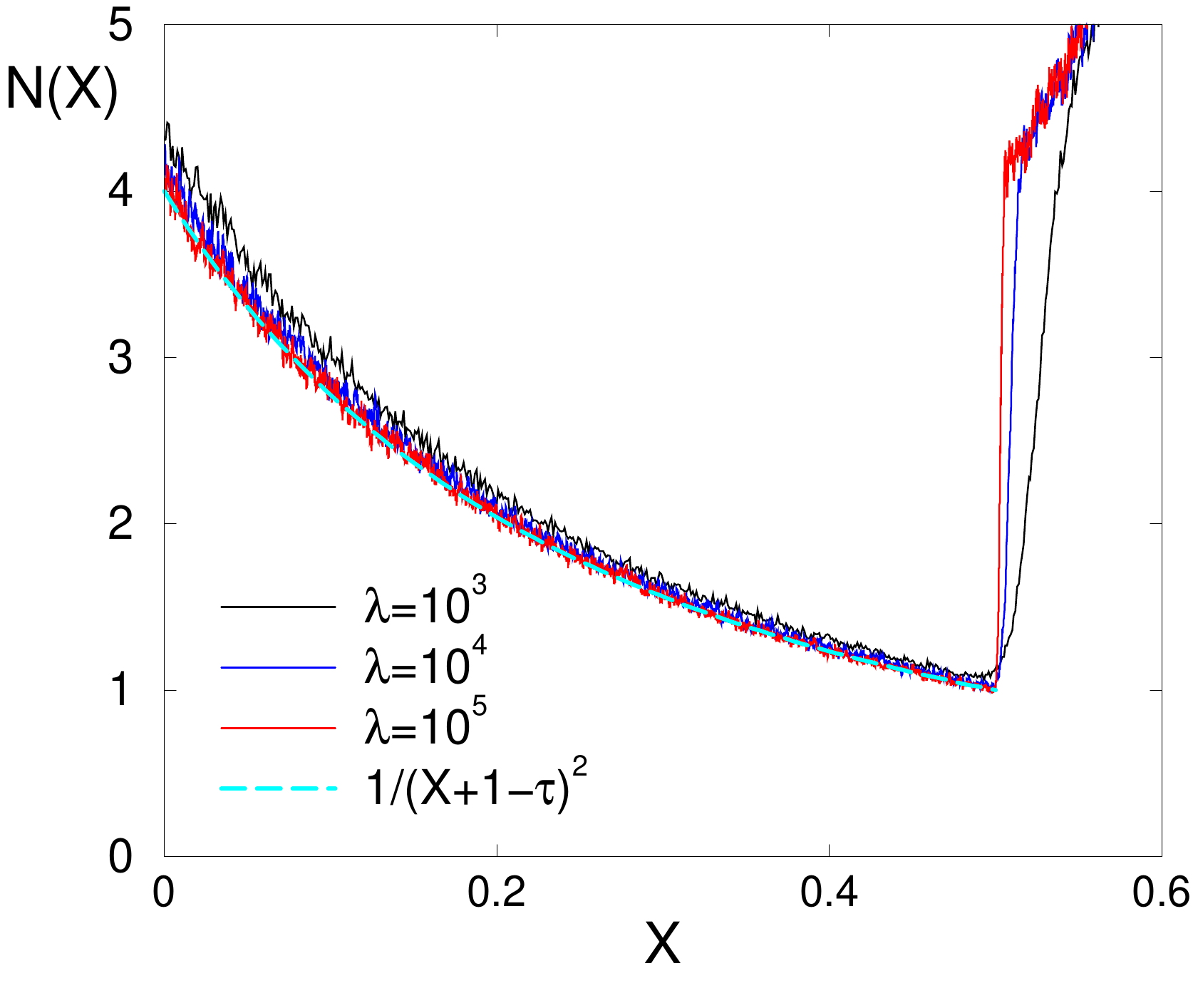}}\qquad\qquad
    \subfigure[]{\includegraphics[width=0.425\textwidth]{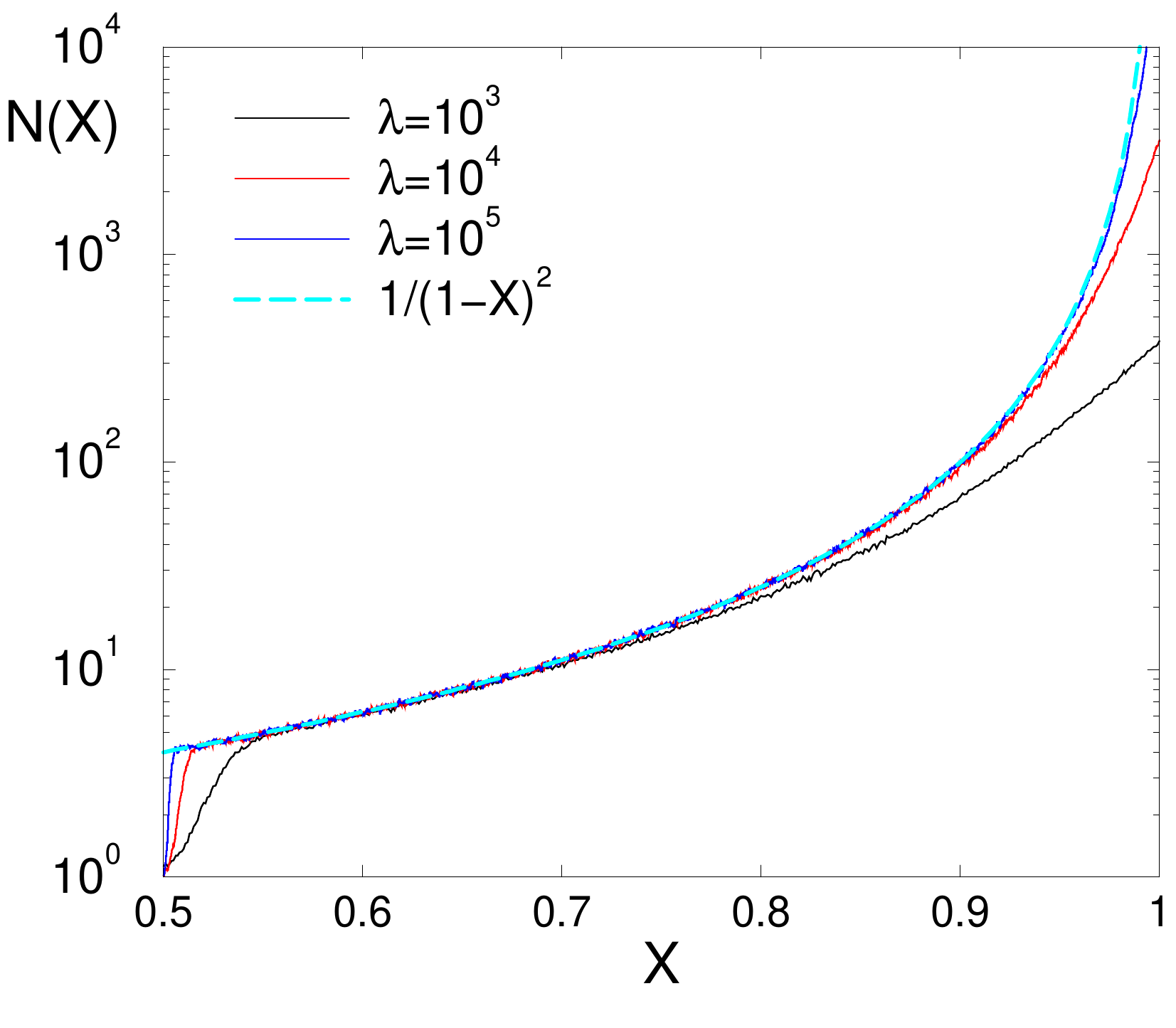}}}
  \caption{The scaled density profile of vacancies for $\tau=\frac{1}{2}$
    in: (a) the active zone $X<\frac{1}{2}$, and (b) the passive zone
    $X>\frac{1}{2}$.  The data for $\lambda=10^4$ and $\lambda=10^5$ are
    smoothed over a 10-point and 100-point range, respectively.}
\label{fig:n-vs-x}  
\end{figure}

Because the lot is nearly full for large $\lambda$, it is more revealing to
focus on the spatial distribution of vacancies $1-\rho(k)$.  Since the
vacancy density is of the order of $1/\lambda$ in the nearly fully occupied
region of the parking lot, it is useful to define the scaled location
$X\equiv k/\lambda$ and the scaled vacancy density
\begin{equation}
N(X)= \lambda[1-\rho(k)] \,.
\end{equation}
With this rescaling, we numerically observe that the vacancy density $N_a(X)$
in the active zone (defined by $X<\tau$ in scaled units) is well fit by the
simple form (Fig.~\ref{fig:n-vs-x}(a)):
\begin{equation}
\label{NX-ansatz}
N_a(X)=\frac{1}{(X+1-\tau)^2} \qquad\qquad 0\leq X\leq \tau\,.
\end{equation}
This conjectured form of the density profile implies that the average number of vacancies
in the active zone is
\begin{align}
  \langle n\rangle_a = \int_0^\tau dX\,N_a(X) = \frac{\tau}{1-\tau}\,.
\end{align}
This result turns out to be asymptotically exact, and we derive it in the
next section without relying on \eqref{NX-ansatz}.

If there are no vacancies in the active zone, the driver must backtrack and
park in the passive zone.  Qualitatively, the parking mechanism in the
passive zone is the mirror image of parking in the active zone.  Based on
this insight, as well as on the numerical data itself, we make the guess that
the scaled vacancy density in the passive zone $\tau<X<1$ is given by
\begin{equation}
\label{NX-front}
N_p(X)=\frac{1}{(1-X)^2} \qquad\qquad \tau < X < 1\,.
\end{equation}
This simple form also provides an excellent fit to the data in the spatial
portion of the parking lot that is nearly full (Fig.~\ref{fig:n-vs-x}(b)).

The jump in the vacancy density profile at $X=\tau$ reflects the nature of
the threshold strategy.  Since a driver starts looking for a vacancy only
when $X=\tau$ is reached, parking spots just closer than $X=\tau$ are likely
to be taken, while the best spots close to $X=0$ are likely to be ``wasted''.
Conversely, if one needs to backtrack to park, it is the spots just beyond
$X=\tau$ that will be taken, while more distant parking spots will remain
more plentiful.  A curious and unexplained feature of the density profiles
\eqref{NX-ansatz} and \eqref{NX-front} is that $N_a(X=0)=N_p(X=\tau)$.

From the density \eqref{NX-front}, we estimate
the number of vacancies in the passive zone by
\begin{align}
\label{np}
  \langle n\rangle_p = \int_\tau^{1-\epsilon} dX\,N_p(X)\,.
\end{align}
Here, we introduce an upper cutoff to the integral because the vacancy
density \eqref{NX-front} cannot hold all the way to $X=1$.  Since the density
profile has a transition zone whose width is of the order of $\lambda^{1/2}$,
the cutoff in scaled units should be $\epsilon\sim \lambda^{-1/2}$.  With
this cutoff, \eqref{np} gives $\langle n\rangle_p\sim \lambda^{1/2}$.  Thus
for large $\lambda$ and with drivers all following the same threshold
strategy, there are of the order of $\lambda^{1/2}$ lousy parking spots
more distant than the threshold, and a few good parking spots
within the threshold.

\section{Number of Vacancies in the Active Zone}
\label{sec:der}

We now derive the distribution of the number of vacancies in the active zone,
from which we deduce the optimality of the $\frac{1}{2}$ rule.  Denote by
$P_n(\tau)$ the probability to find $n$ vacancies in the active zone of the
parking lot, $0<X<\tau$, when the threshold is $\tau$.  Further, define
$V_1(X_1;\tau)$ as the probability density for a single vacancy at $X_1$ and
generally $V_n(X_1,\ldots,X_n;\tau)$ as the probability density for $n$
vacancies at $0<X_1<X_2<\ldots<X_n<\tau$.  The probability to find $n$
vacancies in the active zone is therefore
\begin{equation}
\label{PnV}
P_n = \int \cdots\int_{0<X_1<\ldots<X_n<\tau} dX_1\ldots dX_n\,V_n(X_1,\ldots,X_n;\tau)\,.
\end{equation}
Hereinafter we typically drop the dependence on $\tau$ to declutter the
notation.  Explicitly
\begin{subequations}
\begin{align}
\label{P1}
P_1& = \int_0^\tau dX_1\,\,V_1(X_1)\,,\\
\label{P2}
P_2& = \int_0^\tau dX_1  \int_{X_1}^\tau dX_2\,\,V_2(X_1,X_2)\,,\\
\label{P3}
P_3& = \int_0^\tau dX_1  \int_{X_1}^\tau dX_2 \int_{X_2}^\tau dX_3\,\,V_3(X_1,X_2,X_3)\,,
\end{align}
\end{subequations}
etc.  For notational consistency, we also write $V_0=P_0$ for the probability
of no vacancies in the active zone.

To compute the probability distribution $P_n$ for the number of vacancies in
the active zone, we ostensibly need the densities $V_n$.  As we now show, our
approach circumvents the need for the complete information about these
densities.  The quantity $V_0$ satisfies the rate equation
\begin{equation}
\label{V0:rate}
\frac{dV_0}{dt} = - \lambda \tau V_0+\lambda \int_0^\tau dX_1\,V_1(X_1)\,.
\end{equation}
The loss term on the right accounts for the departure of a car from the fully
occupied active zone.  The gain term accounts for a car that parks in the
single vacancy within the active zone.  In the steady state, \eqref{V0:rate}
becomes
\begin{equation}
\label{V0:inf}
\int_0^\tau dX_1\,V_1(X_1) = \tau V_0\,.
\end{equation}
The left-hand side of \eqref{V0:inf} equals $P_1$ (see \eqref{P1}), so we have
\begin{equation}
\label{P1:sol}
P_1 = \tau V_0\,.
\end{equation}

Similarly to \eqref{V0:rate}, we write the rate equation for $V_1(X_1)$.  In
the steady state, this equation reduces to
\begin{equation}
\label{V1:inf}
\int_{X_1}^\tau dX_2\,V_2(X_1,X_2) +V_0 = \tau V_1(X_1) + V_1(X_1) \,.
\end{equation}
The first term on the left side accounts for parking when two vacancies
exist, while the second term accounts for the departure of a car from the fully
occupied active zone.  The first term on the right accounts for the departure
of a car when the active zone contains a single vacancy, while the second
term accounts for a car that parks at $X_1$.  We now integrate \eqref{V1:inf}
over the active zone, $0<X_1<\tau$.  Using \eqref{P2}, the left-hand side
becomes $P_2+\tau V_0$, while the right-hand side becomes $(1+\tau)P_1$,
after recalling \eqref{P1}. Hence $P_2=(1+\tau)P_1-\tau V_0$, which, in
conjunction with \eqref{P1:sol}, yields
\begin{equation}
\label{P2:sol}
P_2 = \tau^2 V_0\,.
\end{equation}

\begin{figure}[ht]
  \centerline{\includegraphics[width=0.45\textwidth]{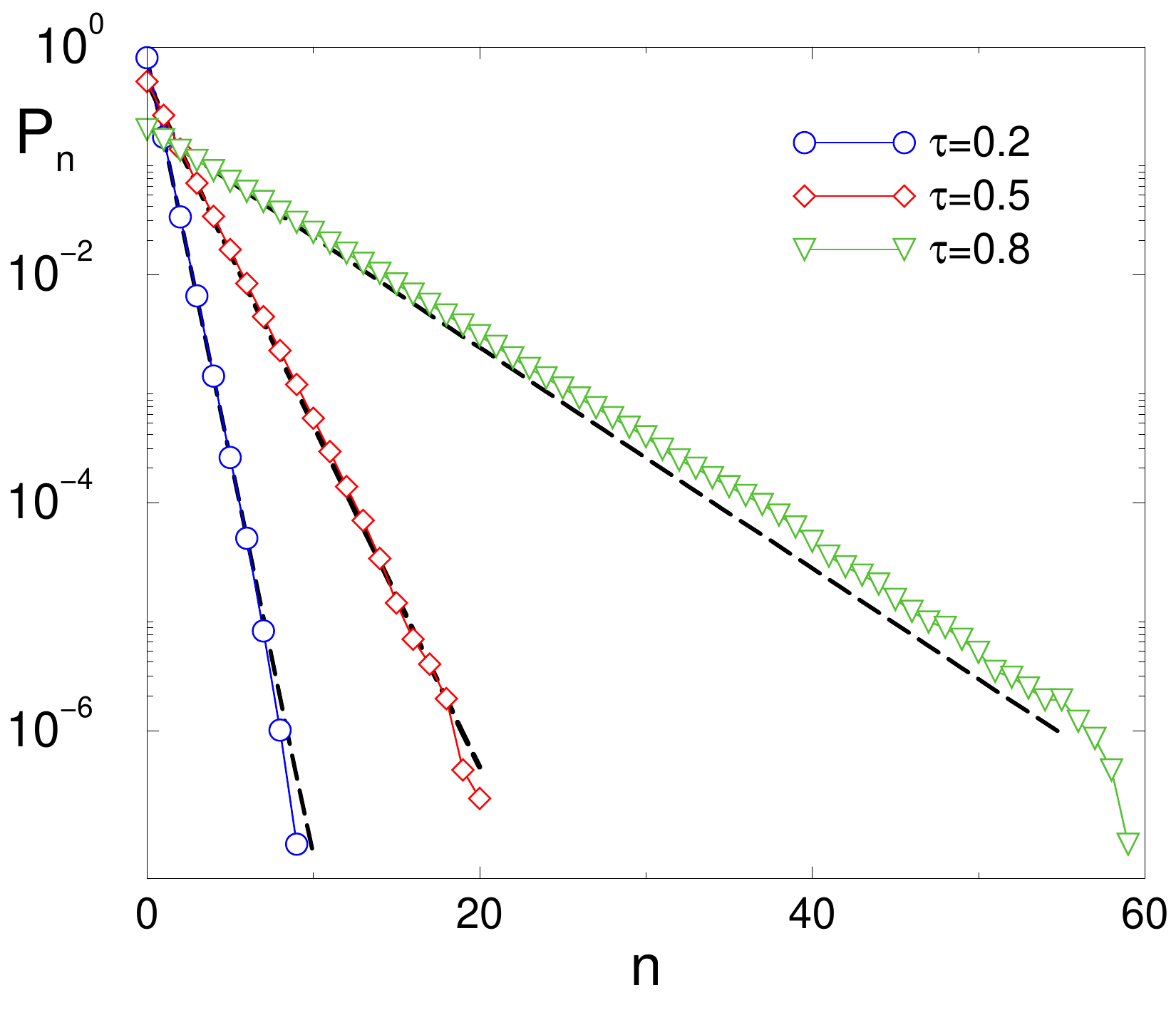}}
  \caption{The probability $P_n$ for $n$ vacancies in the active zone for
    three threshold values and for $\lambda=10^4$.  The dashed lines are the
    theoretical predictions from \eqref{Pk:sol}.}
\label{fig:Pn}  
\end{figure}

Continuing this line of reasoning, we write the rate equation for
$V_2(X_1,X_2)$.  In the steady state, this equation reduces to
\begin{align}
\label{V2:inf}
(\tau+1)V_2(X_1,X_2) &= \int_{X_2}^\tau dX_3\,V_3(X_1,X_2, X_3) + V_1(X_1) + V_1(X_2) \,.
\end{align}
Integrating \eqref{V2:inf} over $0<X_1<X_2<\tau$ ultimately yields
\begin{equation}
\label{P3:sol}
P_3 = \tau^3 V_0\,.
\end{equation}
Following this approach for general $n$, we find $P_n = \tau^n V_0$. The
normalization condition, $\sum_{n\geq 0} P_n=1$, fixes $V_0$ to be $1-\tau$.
Thus the distribution of the number of vacancies in the active zone is the geometric distribution 
\begin{equation}
\label{Pk:sol}
P_n(\tau) = (1-\tau) \tau^n\,.
\end{equation}
The average number of vacancies in the active zone, $0<X<\tau$,
is therefore
\begin{equation}
\label{k:av}
\langle n\rangle_a = \sum_{n\geq 0} nP_n(\tau) = \frac{\tau}{1-\tau}\,.
\end{equation}
Our simulation data in Fig.~\ref{fig:Pn} is in excellent agreement with the
prediction~\eqref{Pk:sol}.

From \eqref{Pk:sol}, the probability to have one vacancy in the active zone
is $P_1(\tau)=\tau(1-\tau)$.  By its definition, this quantity coincides with
the probability that a newly entering car parks in the best available spot.
This probability is maximized when $\tau=\tau_*=\frac{1}{2}$.  At this
optimal threshold value, there is one vacancy, on average, in the active zone
and the probability of actually parking in the best available spot is
$P_1(\tau_*)=\frac{1}{4}$.

\section{The Best and the Actual Parking Spot}
\label{sec:position}

\begin{figure}[ht]
\centerline{\includegraphics[width=0.45\textwidth]{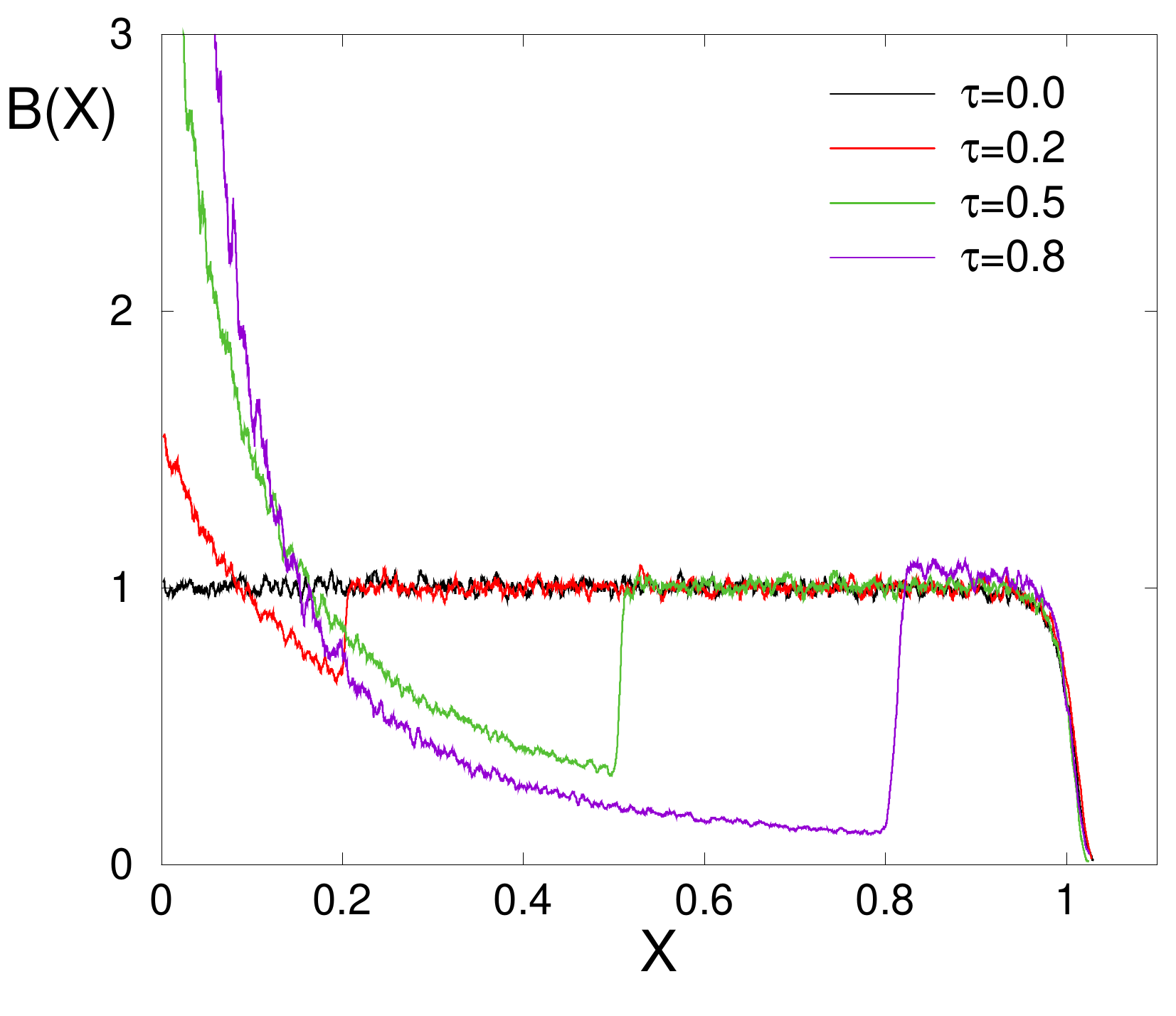}}
\caption{The probability density $B(X)$ that the best parking spot is at a
  scaled distance $X$ from the target for threshold strategies with
  $\tau=0, ~0.2, ~0.5, ~0.8$.  The data are based on $\lambda=10^4$ and are
  smoothed over a 50-point range. }
\label{fig:BX}  
\end{figure}

A natural question that arises upon entering a crowded parking lot is: where
is the best open parking spot?  Let $B(X)$ be the probability density to have
the best parking spot at $X$.  This probability density has a non-trivial
spatial structure (Fig.~\ref{fig:BX}).  For $\tau=0$ (the optimistic strategy
in Ref.~\cite{KR}), the distribution is flat: $B(X)=1$.  That is, if one
drives straight to the target and then starts looking for a parking spot by
backtracking, the closest parking spot is equiprobably anywhere in the lot.

For the general situation, $0<\tau<1$, the distribution $B(X)$ exhibits
different behaviors in the active and passive zones, with a jump at $X=\tau$.
In the passive zone, which gets occupied due to backtracking, the
distribution is again flat; moreover, $B_p(X)=1$ for $X>\tau$.  In the active
zone, $X<\tau$, the probability density $B_a(X)$ is a decreasing function of
$X$ which becomes progressively more peaked as $\tau$ increases.  That is, if
one starts looking for parking spots when $X$ reaches $\tau$, it is the
parking spots with $X\ltwid \tau$ that are more likely to be filled, while
the desirable spots near $X=0$ will be relatively unfilled.  The probability
density to have the best spot next to the target, $B_a(0)$, is obviously
equal to the vacancy density $N_a(0)$; this implies that
$B_a(0)=(1-\tau)^{-2}$.  The probability density $B(X)$ is normalized, so
that $\int_0^\tau dX\,B_a(X)=\tau$. These are two exact properties of the
probability density $B_a(X)$ in the active zone.  The challenge is to
determine $B_a(X)$ analytically.

The threshold-$\frac{1}{2}$ rule leads to success, viz., to parking in the
best available spot, in a fraction $P_1(\frac{1}{2})=\frac{1}{4}$ of
attempts; otherwise parking occurs in a suboptimal spot.  Thus a useful
measure of the efficacy of the threshold strategy is the ratio
$r\equiv X/X_B$ of the \emph{actual} parking location $X$ to the best
available location $X_B$.  Denote by $q(r)$ the probability density that
$X/X_B=r$.  We know that $r=1$ if there is either 1 or 0 vacancies in the
active zone. In these two cases, the car parks in the best available spot
without backtracking or with backtracking, respectively. The two events occur
with probability $P_0+P_1=1-\tau^2$.  Hence
\begin{equation}
q(r)=(1-\tau^2)\delta(r-1)+Q(r).
\end{equation}
The continuous part $Q(r)$ of the probability density accounts for the
situations when there is more than one vacancy in the active zone, so the
entering car necessarily parks in a suboptimal spot and $r>1$.

\begin{figure}[ht]
  \centerline{\includegraphics[width=0.45\textwidth]{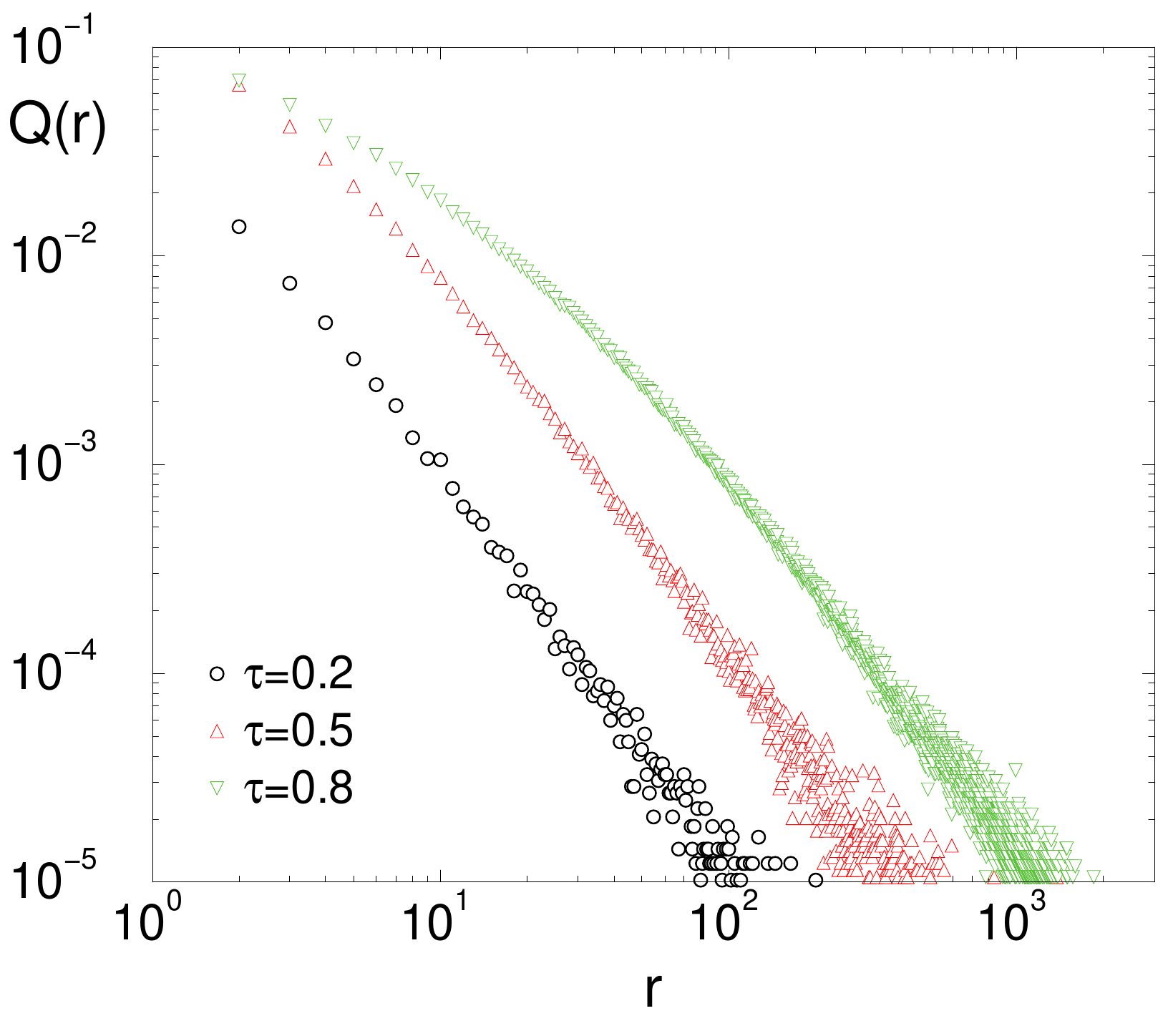}}
  \caption{The continuous part $Q(r)$ of the probability density of the ratio of the actual
    parking location to the best parking location, $r=X/X_B$.}
\label{fig:ratios}  
\end{figure}

The continuous part $Q(r)$ of the probability density satisfies
$\int_1^\infty dr\,Q(r)=\tau^2$ and it appears to have a power-law tail for
large $r$ (Fig.~\ref{fig:ratios}) in the general situation when $0 <\tau <1$.
For the threshold values shown in this figure, the data is reasonably fit by
the power law $r^{-\nu}$, with $\nu\approx 1.8$.  That is, even though a
driver typically does not park in the best spot, it is likely that parking
occurs close to the best spot.  When $\tau=0$ (the optimistic strategy in
\cite{KR}), the car necessarily parks in the best spot by backtracking:
$q(r)=\delta(r-1)$.


\begin{figure}[ht]
  \centerline{\includegraphics[width=0.45\textwidth]{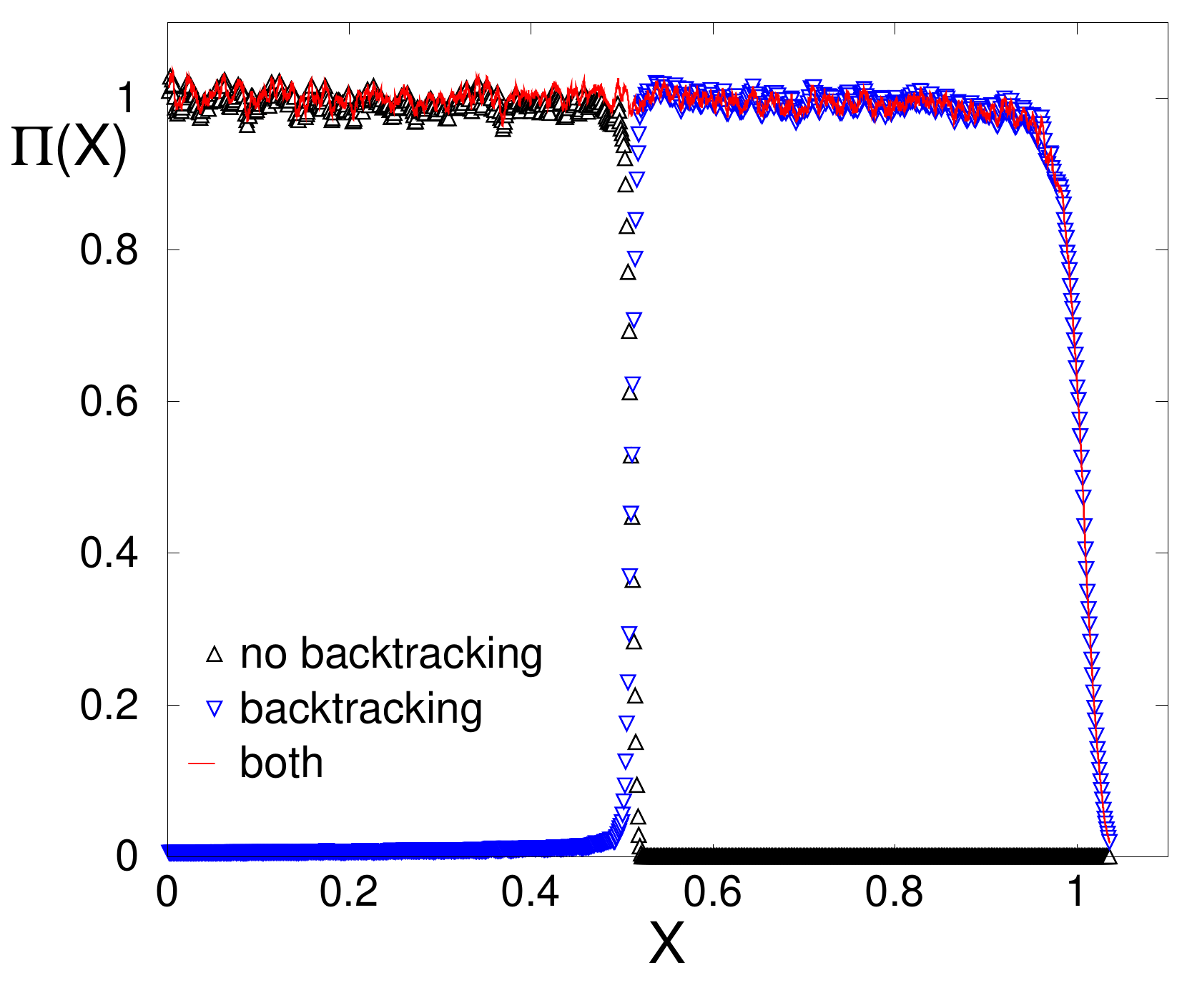}}
  \caption{The spatial distribution of parking locations for
    $\tau=\frac{1}{2}$.  The data have been smoothed over a 50-point range
    and only every tenth data point is shown.}
\label{fig:actual}  
\end{figure}

Although the ratio of the actual parking location to the best location has a
non-trivial distribution, the spatial distribution of the actual parking
location $\Pi(X)$ is remarkably simple: $\Pi(X)=1$ (Fig.~\ref{fig:actual}).
To understand this striking result, consider the part of this distribution
$\Pi_a(X)$ in the active zone, $X\leq \tau$, when a car finds a parking spot
without backtracking.  The strategy in the active zone is effectively the
same as the strategy with an effective threshold $\tau_\text{eff}=1$ in the
case where the span is $\tau L$.  The properly scaled spatial variable is
$X/\tau$ and hence
\begin{subequations}
\begin{equation}
\label{Pi-active}
\Pi_a(X) = \Phi(y), \quad y = \frac{X}{\tau}\,.
\end{equation}

When backtracking occurs, cars enter the passive zone and park in the first
gap, so effectively they follow the prudent strategy with span $(1-\tau) L$.
The properly scaled spatial variable is now $(1-X)/(1-\tau)$ as the cars move
to right when they backtrack.  Thus the part of the parking distribution
$\Pi_p(X)$ that corresponds to the passive zone is
\begin{equation}
\label{Pi-passive}
\Pi_p(X) = \Phi(y), \quad y = \frac{1-X}{1-\tau}\,.
\end{equation}
\end{subequations}
The fractions of cars that park in the active and passive zones are $\tau$
and $1-\tau$.  These fractions also equal $\int_0^\tau dX\,\Pi_a(X)$ and
$\int_\tau^1 dX\,\Pi_p(X)$, respectively.  Using
\eqref{Pi-active}--\eqref{Pi-passive} we indeed recover $\tau$ and $1-\tau$
after imposing the normalization condition
\begin{equation}
\label{Phi}
\int_0^1 dy\,\Phi(y)=1\,.
\end{equation}

To determine the form of the probability density $\Pi(X)$, let us compute its
derivatives at the threshold location $X=\tau$:
\begin{align*}
\frac{d^n \Pi}{d X^n}\Big|_{X=\tau-0}&=\tau^{-n}\Phi^{(n)}(1)\,,\\[2mm]
\frac{d^n \Pi}{d X^n}\Big|_{X=\tau+0}&=(-1)^n (1-\tau)^{-n}\Phi^{(n)}(1)\,,
\end{align*}
where $\Phi^{(n)}(y)=d^n \Phi(y)/dy^n$. The derivatives from left and write
coincide only if $\Phi^{(n)}(1)=0$ for all $n\geq 1$.  Thus $\Pi(X)$ is
smooth at $X=\tau$ when all derivatives of $\Phi(y)$ vanish at $y=1$.  The
vanishing of all derivatives implies\footnote[1]{We tacitly assume
  that $\Phi(y)$ is analytic, namely its Taylor series (centered at $y=1$)
  converges. There are infinitely differentiable functions which are not
  analytic; for such function, vanishing of all derivatives does not assure
  that the function is constant.} that the distribution is uniform, and the
actual value $\Phi(y)=1$ is fixed by the normalization condition~\eqref{Phi}.
Our above argument is not a proof, but it shows that \emph{postulating}
analyticity of the probability density $\Pi(X)$ at the threshold $X=\tau$
suffices to derive its uniformity.

\section{Cost of Parking}
\label{sec:cost}

Maximizing the probability to park in the best available spot is natural and
may be compatible with the intrinsic irrationality of human behavior in
parking lots (see, e.g., \cite{traffic08}).  However, the threshold parking
rule is not necessarily the most rational.  A more logical approach is to
minimize the cost of parking, which we define as the arrival time to reach
the target.  This arrival time is the sum of the time spent in driving in the
lot to find the parking spot plus the walking time from the parking spot to
the target.  In appropriate units where the walking speed equals 1, the
walking time is just the distance from the parking spot to the target, while
the driving time is the driving distance in the lot divided by the driving
speed.  This driving time also equals the driving distance multiplied by the
ratio of the walking speed to the driving speed in the lot; we denote this
ratio by $\epsilon$.

For the $\tau$-threshold parking strategy, these arrival times are
\begin{equation*}
  \frac{\tau}{2}+\epsilon\left(1-\frac{\tau}{2}\right) \qquad\text{and}\qquad
  \frac{1+\tau}{2}+\epsilon\left(1+\frac{1+\tau}{2}\right) \,,
\end{equation*}
for the cases of parking in the active and passive zones, respectively.  Here
we have made use of the result that the parking location is uniformly
distributed in both the active and passive zones.  Multiplying these arrival
times by the probabilities for these two types of parking events, $\tau$ and
$1-\tau$ respectively, the average parking cost is
\begin{equation}
  \label{cost}
\tfrac{1}{2}(1+ 3\epsilon) - \epsilon\tau^2\,.
\end{equation}
This cost function attains a minimum with the prudent strategy, $\tau=1$.
Despite being optimal according to the above choice for the cost function,
the prudent strategy is psychologically upsetting to most people.  In the
prudent strategy, the parking spot that a driver takes is essentially never
the best available, as the typical number of vacancies is of order
$\sqrt{\lambda}$.  Perhaps even more upsetting to a driver who has just
parked is that the absolutely best parking spot---the one that is adjacent to
the target---will be available in approximately $89\%$ of all
realizations~\cite{KR}.

In deriving \eqref{cost}, we tacitly assumed that $\epsilon>0$; however, the
limiting case of $\epsilon=0$ is actually quite natural.  Setting
$\epsilon=0$ is equivalent to only counting the time spent in walking to the
target in the cost of parking.  Since moving in a car requires little
physical effort, drivers with limited walking ability, as well as drivers
with little children, may find this cost function more suitable.
Furthermore, if the target is a store were the drivers tend to buy heavy
goods, it is again reasonable to minimize the time of walking from the target
to the parked car, i.e., set $\epsilon=0$.  In this case, the average parking
cost is independent of strategy.  Since the $\frac{1}{2}$ rule maximizes the
probability of snagging the best parking spot, which everyone prefers, it is
rational to adopt the threshold-$\frac{1}{2}$ rule if the cost of walking
through the parking lot is prohibitively high.

\section{Discussion} 
\label{sec:disc}

We investigated a class of threshold parking strategies in which a driver who
enters a parking lot ignores all open spots a distance greater than $\tau L$
from the target.  Starting at a distance $\tau L$, which defines the end of
the active zone, the driver parks at near end of the first gap encountered.
Here $L$ is the distance from the target to the last parked car and $\tau$ is
the risk threshold.  If there are no available spots in the desirable active
zone, the driver backtracks and parks in the first spot encountered by
backtracking.

When the ratio of the arrival rate to departure rate of each car is large,
$\lambda\gg 1$, open parking spots are rare in the active zone, and the spots
that do exist are more likely to be close to the target
(Fig.~\ref{fig:rho-vs-x}(b)).  Our primary result is that the probability for
a newly entering car to park at the best available spot is maximized for
$\tau=\frac{1}{2}$.  At this optimal threshold, the probability of parking in
this best spot is $\frac{1}{4}$.  We conjecture that our strategy is the best
amongst all (deterministic), and not necessarily threshold parking
strategies.  Proving or refuting this conjecture represents an appealing
challenge.

Our analysis for number of vacancies in the active zone (Sect.~\ref{sec:der})
tacitly assumed that $0<\tau<1$; however, our main result \eqref{Pk:sol} can
be continued to $\tau=0$ and $\tau=1$; these limits correspond to the
optimistic and prudent strategies in our earlier study~\cite{KR}.  This
continuation correctly predicts that, for the optimistic strategy, the
probability of finding the best spot without backtracking vanishes.  Indeed,
for this strategy, backtracking occurs with probability
$\lambda/(1+\lambda)$~\cite{KR}, so backtracking is asymptotically certain in
this case.  For the prudent strategy, the probability of backtracking
vanishes as $\lambda^{-1}$, but the probability that the spot where the car
actually parks is closest to the target vanishes as $\lambda^{-1/2}$.  Thus
the $\frac{1}{2}$ rule is superior to both the optimistic and prudent
strategies in finding the best available parking spot.

While we found an optimal parking strategy, we were unable to
analytically determine several natural spatial properties of the parked cars,
such as the distribution of the number of vacancies $V$, the number of gaps
(Fig.~\ref{Fig:cartoon-threshold}), the distribution of the span $L$, and
multisite densities, such as $V_n(X_1,\ldots,X_n)$ in the active zone.
Additionally, the expressions~\eqref{NX-ansatz} and \eqref{NX-front} for the
vacancy densities in the active and passive zones, which appear to be
correct, are conjectural.  The decoupling approximation, which we employed
in~\cite{KR} to determine the average density of cars (or vacancies) at a
given spatial location, fails for the threshold strategy and another
analytical approach is needed.

Our threshold parking rules are inspired by
the strategies that arise in optimal stopping problems and in decision theory
(see, e.g., \cite{L61,D70,Chow71,BG84,P99}). Some of the methods developed in
this body of work, particularly in the realm of the famous ``secretary
problem''
\cite{D70,Chow71,BG84,P99,Dynkin,Chow64,Gilbert,F89,Bruss00,Bruss03,Dendievel,GKMN,FW10,GM13},
may prove useful to analyze optimal parking.  Indeed, the best strategy for
various decision theory optimization problems is often a threshold
strategy~\cite{L61,Dynkin}.

An intriguing contribution to the work on optimal stopping problems is the
odds algorithm~\cite{Bruss00}, which is summarized by adage: sum the odds to
one and stop. The odds algorithm has been proved for some optimal stopping
problems with independent events~\cite{Bruss00,Bruss03,Dendievel}.  We now
show that the odds algorithm holds in our case where parking events are not
independent. By definition, the odds are the ratios $p/(1-p)$ where the $p$'s are the
probabilities of success.  In our situation, the probability of successful
parking at spot $k$ is the probability $1-\rho(k)$ that this spot is empty 
\footnote{When a driver meets the first gap in the active zone, parking
  occurs at the spot in the gap that is closest to the target; more distant
  vacancies in the gap are ignored.  However, gaps in the active zone
  typically consists of isolated vacancies. Indeed, the total number of gaps
  with $\ell$ vacancies scales as $\lambda^{-(\ell -1)}$ and hence vanishes
  in the $\lambda\to\infty$ limit for all $\ell\geq 2$.}.  Since the
$\rho(k)$ are very close to 1 in the bulk of the span, the odds coincide with
the probabilities (i.e., the vacancy densities) to leading order. Summing the
vacancy densities in the active zone yields the average number of vacancies,
$\langle n\rangle_a=\tau/(1-\tau)$.  Equating this number to one recovers the
$\frac{1}{2}$ rule.

It is worth pointing out that our model assumes a homogeneous population of
drivers; namely, they all have the same parking threshold parameter $\tau$.
Rich behaviors may emerge for heterogeneous populations.  The simplest
heterogeneous population consists of drivers that each have independent
random parking thresholds that, for example, are uniformly distributed on
$(0,1)$.  There is no longer an optimization problem to be solved, but the
spatial properties of the parked cars could be interesting and tractable.

This heterogeneous model is a sort of equilibrium counterpart of the hashing
problem that was originally introduced by Konheim and Weiss~\cite{KW66} and
described by Knuth~\cite{Knuth-3} using the parking problem language.  This
hashing problem and its numerous extensions have been subsequently studied by
many authors, see, e.g.,~\cite{FPV98,Janson01,Janson05,FS09,Alois16} and
references therein.  In the standard application of the hashing problem, one
attempts to write a fixed-size file onto a computer disk.  A random location
on the disk is picked and if the file can be accommodate there, the file is
written.  If not, the next location is selected.  If this location is empty,
the file is written there.  If not, continue moving in one direction until a
vacancy is encountered.  The list of starting locations is the hash table.
In the language of parking, the lot is finite, there is no departure
mechanism (files are not erased), so eventually the parking lot becomes full.
Richer behaviors arise when the file sizes are variable.  The correspondence
between hashing and parking offers the possibility that the substantial
research literature on the hashing problem could guide developments in our
parking problem.

\medskip
PLK thanks the hospitality of the Santa Fe Institute where this work was
initiated.  SR gratefully acknowledges financial support from NSF grant
DMR-1608211. We also thank John Miller for helpful conversations.

\bigskip
\bigskip\newpage
\newcommand{\newblock}{}

\end{document}